\documentclass[journal,twoside,web]{ieeecolor}
\usepackage{generic}
\usepackage{amsmath,amssymb,amsfonts}
\usepackage{algorithmic}
\usepackage{graphicx}
\usepackage{textcomp}
\usepackage{etoolbox}
\usepackage{xcolor}
\usepackage[colorlinks=true,
            linkcolor=blue,
            citecolor=blue,
            urlcolor=magenta]{hyperref}

\usepackage{amsmath,amssymb,amsfonts}
\usepackage{algorithmic}
\usepackage{graphicx}
\usepackage{textcomp}
\usepackage[switch]{lineno}

\usepackage{cite}

\def\BibTeX{{\rm B\kern-.05em{\sc i\kern-.025em b}\kern-.08em
    T\kern-.1667em\lower.7ex\hbox{E}\kern-.125emX}}
\begin{document}

\title{On Optimizing Electrode Configuration for Wrist-Worn sEMG-Based Thumb Gesture Recognition}

\author{Wenjuan Zhong, Chenfei Ma, and Kianoush Nazarpour, \IEEEmembership{Senior Member, IEEE}
\thanks{This work was funded by Engineering and Physical Sciences Research Council, UK (EP/R004242/2, EP/Y028856/1, EP/X031950/1). K.N. is also supported by the National Institute for Health and Care Research (NIHR) HealthTech Research Centre in long term conditions (Devices for Dignity). \textit{(Corresponding author: Kianoush Nazarpour. (e-mail: kianoush.nazarpour@ed.ac.uk).}}
\thanks{All authors are with School of Informatics, The University of Edinburgh, Edinburgh EH8 9AB, United Kingdom.}}

\maketitle

\begin{abstract}
Thumb gestures provide an effective and unobtrusive input modality for wearable and always-available human–machine interaction. Wrist-worn surface electromyography (sEMG) has emerged as a promising approach for compact and wearable human-machine interfaces. However, compared to forearm sEMG, the impact of electrode configuration on wrist-based decoding performance remains understudied. We systematically investigated electrode configuration strategies for wrist-based thumb-movement recognition using high-density (HD) and low-density (LD) sEMG measurement systems. We considered factors such as muscle region, reference scheme, channel count, and spatial density of the electrode. Experimental results show that 1) extensor-side electrodes outperform flexor-side electrodes (HD: 0.871 vs. 0.821; LD: 0.769 vs. 0.705); 2) monopolar recordings consistently outperform bipolar configurations (15-channel with HD monopolar vs. LD bipolar: 0.885 vs. 0.823); and 3) increasing channel count enhances performance, but exhibits diminishing returns. We further show that electrode spatial distribution introduces a trade-off between spatial coverage and compactness. The findings suggest that the effectiveness of wrist-worn sEMG systems depends less on the deployment of a large number of electrodes in a broad sensing area and more on the optimization of electrode placement and the referencing scheme. This work provides practical guidelines for developing efficient wrist-worn sEMG-based gesture recognition systems.

\end{abstract}

\begin{IEEEkeywords}
Electromyography (EMG), thumb gestures, wrist, electrode configuration.
\end{IEEEkeywords}

\section{Introduction}
\label{sec:introduction}

The thumb plays a central role in human dexterity, contributing nearly 40\% of hand function through its ability to resist and generate controlled forces \cite{emerson1996anatomy, yue2017hand}. Thumb-oriented actions such as pinching, tapping, and directional swiping are fundamental to daily tasks and are increasingly adopted as input in wearable, robotic, and AR/VR systems \cite{kieliba2021robotic, kin2024stmg}. Furthermore, the thumb retains a high degree of kinematic independence even when the hand is engaged in object manipulation and human–machine interfaces \cite{lee2025grab}. Recognizing these thumb movements for spatial computing applications remains a challenge due to their kinematic signatures. Although optical methods have been explored for thumb gesture recognition \cite{kim2012digits, kin2024stmg}; such approaches are constrained by occlusion, power consumption, and limited usability for mobile wearable applications.

Surface electromyography (sEMG) provides direct access to the underlying neuromuscular activation and enables the decoding of motion intent even when external kinematic and kinetic cues are weak. Consequently, most existing sEMG-based gesture recognition systems are developed to decode multi-finger movements and wrist movements, which typically place electrodes on the forearm to capture muscle activity from larger muscle groups \cite{zhong2025deep, du2017surface, jiang2021open, ma2024distanet, jiang2021emerging}. For thumb movement recognition, wrist-based sEMG sensing is preferred for two reasons. First, thumb-related muscles are more spatially concentrated on the surface of the wrist, whereas toward the forearm, they are located deeper and overlap with other muscles, increasing crosstalk in sEMG recordings. Secondly, compared to additional forearm-mounted sensing devices, integrating sEMG-based interaction on wrist-worn wearable devices, such as smartwatches and fitness bands, enables more natural deployment and supports long-term comfortable daily use \cite{cao2024finger}.

In recent years, increasing attention has been directed towards wrist-based sEMG gesture recognition using low-density (LD) bipolar electrodes \cite{jiang2017feasibility, botros2020electromyography, meredith2024comparing, botros2025zero, he2025improving, simpetru2025myogestic} and high-density (HD) monopolar electrode arrays \cite{guerra2022far, yang2025non}. Using LD configurations, Botros \textit{et al.} \cite{botros2020electromyography} compared sEMG signals recorded at the wrist and forearm with four bipolar electrodes during multi-finger and wrist movements, and reported that wrist sEMG exhibited higher signal quality and superior performance for gestures involving fine finger control. He \textit{et al.} \cite{he2025improving} proposed a multi-position training strategy to mitigate performance degradation caused by variations in limb-condition in wrist sEMG recordings, also using four bipolar electrodes. Beyond LD sensing, several studies have explored wrist sEMG using HD monopolar arrays with signal decomposition techniques. Guerra \textit{et al.} \cite{guerra2022far} investigated the firing characteristics of motor neurons using blind source separation and demonstrated the feasibility of gesture recognition based on decoded motor unit activity. Similarly, Yang \textit{et al.} \cite{yang2025non} reported a strong correlation between decoded motor unit activity at the wrist and forearm, and successfully classified finger gestures in tetraplegic individuals using motor neuron potentials obtained from wrist and forearm recording sites.

\begin{figure*}[ht]
\centerline{\includegraphics[width=\textwidth]{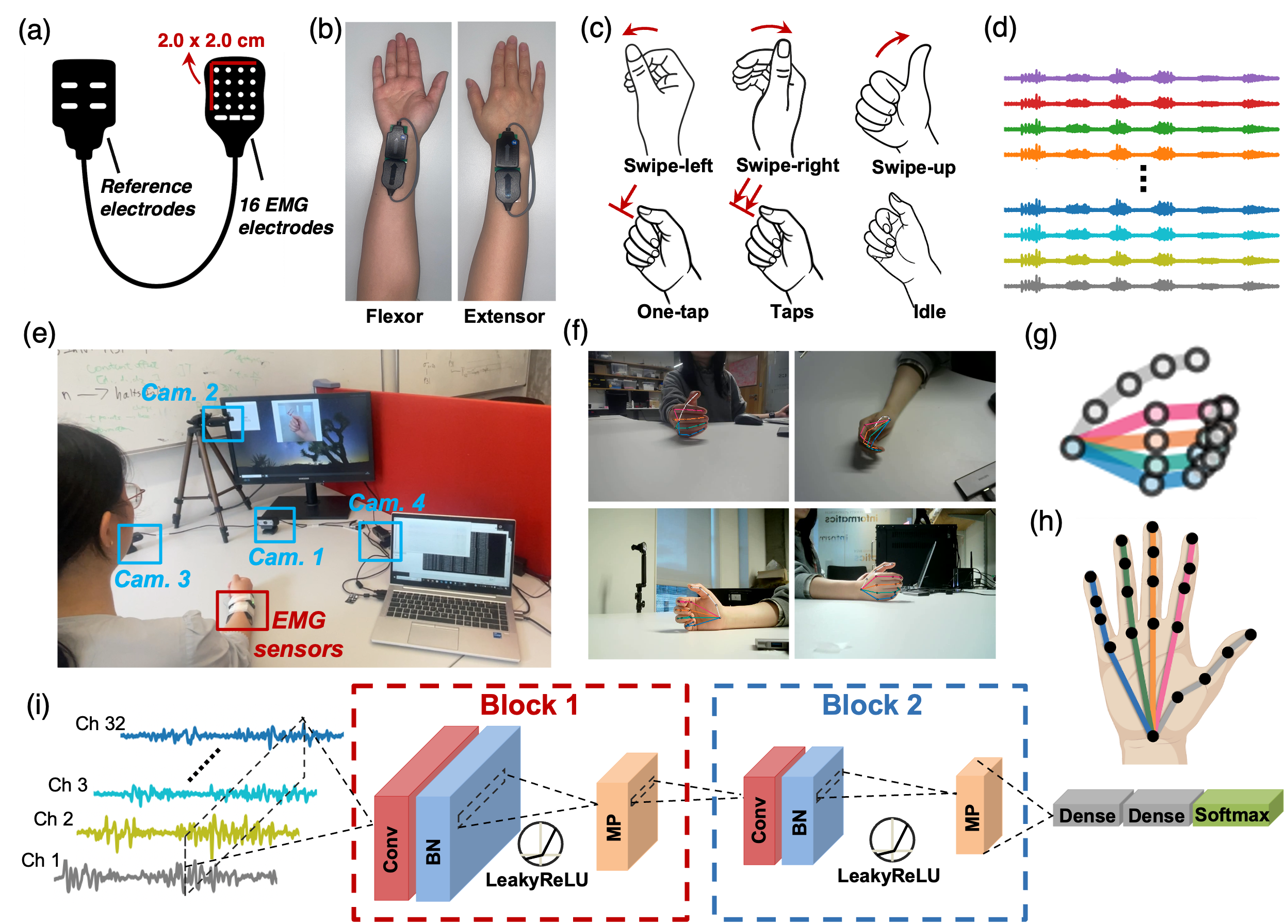}}
\caption{Overview of proposed study illustrating the experimental setup and the deep learning framework for thumb gesture recognition. (a) Trigno Maize sensor, a high-density sEMG electrode grid including 16 channels. (b) Placement of two sensor grids on the extensor and flexor sides of the right wrist, with the central column aligned with the forearm midline. (c) Illustration of the thumb movements. (d) Example of recorded sEMG signals. (e) Experimental setup showing synchronized sEMG acquisition and multi-camera recording. (f) Camera views used for hand motion capture. (g) Reconstructed three-dimensional hand kinematics obtained from multi-view camera recordings. (h) Visualization of hand joint landmarks. (i) CNN architecture. }
\label{fig1_system}
\end{figure*}

Although prior work suggests that wrist and forearm sEMG can achieve comparable signal quality regardless of the use of LD or HD sensors \cite{ botros2020electromyography, yang2025non}, sEMG acquisition on the wrist is subject to fundamentally different constraints. First, the detection area available on the wrist for electrode placement is smaller. Second, extrinsic muscle activation patterns for movements that are performed only by the thumb are spatially compressed at the wrist. However, it remains unclear how wrist-based sEMG electrode configurations influence the performance of thumb gesture recognition, including factors such as wrist muscle regions, reference scheme (monopolar vs. bipolar), channel counts, and spatial density of the electrodes. 

In this study, we systematically investigate electrode configuration strategies for wrist-based thumb gesture recognition using sEMG signals. To this end, we employ a convolutional neural network (CNN) and a temporal convolutional network (TCN) as representative deep learning models for sEMG-based feature extraction \cite{zhong2025deep}. We evaluate decoding performance for thumb gesture recognition under a range of electrode configuration factors, including muscle region selection, channel count, and spatial density, and compare monopolar and bipolar recording schemes using both high-density (HD) electrode grids and low-density (LD) bipolar electrodes. This study aims to provide a systematic analysis of how electrode configuration affects decoding performance and to derive design guideline for wrist-worn sEMG systems.

\section{Methodology}
\subsection{Experimental Setup and Participants}
A multi-modal experimental platform was developed to enable the synchronized acquisition of sEMG signals and hand movements using a multi-camera system, as illustrated in Fig.~\ref{fig1_system}. Ten able-bodied participants (four males and six females), all right-handed, were recruited for the study. All participants signed an informed consent form approved by the local ethics committee at the University of Edinburgh (Reference Number: 2019/89177), in accordance with the Declaration of Helsinki.

\subsubsection{sEMG Acquisition} sEMG signals were acquired using two HD electrode grids (Trigno Maize sensor, Delsys Inc., USA), each consisting of a $4 \times 4$ array with 5~mm inter-contact spacing, as shown in Fig.~\ref{fig1_system}(a). Two grids were placed on the extensor and flexor sides of the right hand wrist, with the central column aligned with the midline of the forearm (Fig.~\ref{fig1_system}(b)).

Before electrode placement, the skin was washed with an alcoholic solution. 32 monopolar sEMG signal channels (Fig.~\ref{fig1_system}(d)) were recorded using the wireless Trigno Maize system (Delsys Inc., USA). The signals were band-pass filtered (20–450~Hz), prior to offline analysis, sampled at 1000~Hz, and transmitted to a host computer via a USB port. Data acquisition was performed using a customized graphical user interface built on the Delsys API (AeroPy Layer). 

In addition to HD recordings, bipolar LD sEMG signals were collected using the Trigno Quattro sensor (Delsys Inc., USA). 15 bipolar channels were evenly distributed around the wrist. Detailed placement and configuration of the Quattro sensors are provided in the \textit{Supplement Document}.

\subsubsection{Hand Movements Recording using Cameras} 

During sEMG acquisition, synchronized multi-view video recordings were captured with four cameras at a resolution of $680 \times 480$ pixels and a frame rate of 30~Hz (Fig.~\ref{fig1_system}(e)). The cameras were positioned around the experimental workspace (front, upper, left, and right) to capture different views of hand movement (Fig.~\ref{fig1_system}(f)).

The recorded videos were processed using a pre-trained neural network from the MediaPipe framework~\cite{lugaresi2019mediapipe} to extract two-dimensional (2D) coordinates of 21 joint landmarks of the hand (Fig.~\ref{fig1_system}(h)). The key points were then triangulated into three-dimensional (3D) coordinates (Fig.~\ref{fig1_system}(g)) using the Anipose library \cite{karashchuk2021anipose}.

\subsection{Experimental Protocols}
Six thumb gestures were included in the experiments: \textit{idle}, \textit{swipe-left}, \textit{swipe-right}, \textit{swipe-up}, \textit{one-tap}, and \textit{taps}, as illustrated in Fig.~\ref{fig1_system}(c). The \textit{idle} gesture is static, while the remaining five thumb gestures were dynamic movements. Each trial lasted 4~s. For \textit{swipe-left}, \textit{swipe-right}, \textit{swipe-up}, and \textit{one-tap} gestures, the motion consisted of four phases, including movement ($\sim$0.5~s), hold ($\sim$1.5~s), return ($\sim$0.5~s), and idle ($\sim$1.5~s). The details of the thumb movements are as follows:

\begin{itemize}
    \item \textbf{Idle}: The thumb remains relaxed and static in a neutral resting position.
    \item \textbf{Swipe-left}: The thumb moves leftward from the neutral position, holds briefly at the target position, and then returns to the neutral position before relaxing.
    \item \textbf{Swipe-right}: The thumb moves rightward from the neutral position, holds briefly at the target position, and then returns to the neutral position before relaxing.
    \item \textbf{Swipe-up}: The thumb moves upward from the neutral position, holds briefly at the target position, and then returns to the neutral position before relaxing.
    \item \textbf{One-tap}: The thumb moves downward to tap the index finger, holds briefly at the contact position, and then returns to the neutral position before relaxing.
    \item \textbf{Taps}: The thumb repeatedly taps the index finger for a duration of 2~s, and then returns to the neutral position before relaxing.
\end{itemize}

Each participant completed two recording sessions: one using the Maize sensor and one using the Quattro sensor. The sessions were separated by a 30-minute rest period to reduce fatigue effects. To counterbalance potential order effects, five participants performed the Maize session first, while the remaining five began with the Quattro session. To minimize placement-related variability, both sensors were positioned in the same spot throughout the sessions.

Each session consisted of ten blocks. Each block contained 31 trials (31 $\times$ 4~s = 124~s), including an initial \textit{swipe-up} trial for participant preparation, followed by six repetitions of each of five dynamic gestures presented in random order (6 $\times$ 5). Visual cues depicting the target gesture were displayed on a monitor, and auditory cues were provided to indicate the start and return phases of the movement. The participants rested for about 2 minutes between blocks.

\subsection{sEMG Data Processing and Windowing}
Raw sEMG signals were filtered using a fourth-order Butterworth filter between 20 and 450~Hz. The initial \textit{swipe-up} trial was excluded for postprocessing. For the remaining 30 trials, a dynamic thumb-movement segment from 0.5 to 2.0~s was extracted to avoid reaction-time effects and onset transients in each trial, which was labeled according to the corresponding gesture class. A segment from 3.4 to 3.9~s was extracted and used to represent the idle class.

The processed sEMG signals were segmented into 250 ms windows with 50\% overlap, which served as input to the gesture decoding model. The length of the segmented window directly determines the temporal resolution of the sEMG representation and the amount of neuromuscular information captured within each window \cite{khushaba2021decoding, lin2025leveraging}. A window length of 250~ms has been widely adopted in gesture recognition based on sEMG, as it provides an effective trade-off between capturing discriminative muscle activation patterns and maintaining low computational cost \cite{junior2023semg, prakash2025optimized}. The overlap of 50\% is commonly used to preserve the temporal continuity between adjacent windows and reduce information loss at the window boundaries \cite{li2024optimizing}.

\begin{figure}[!t]
\centerline{\includegraphics[width=\columnwidth]{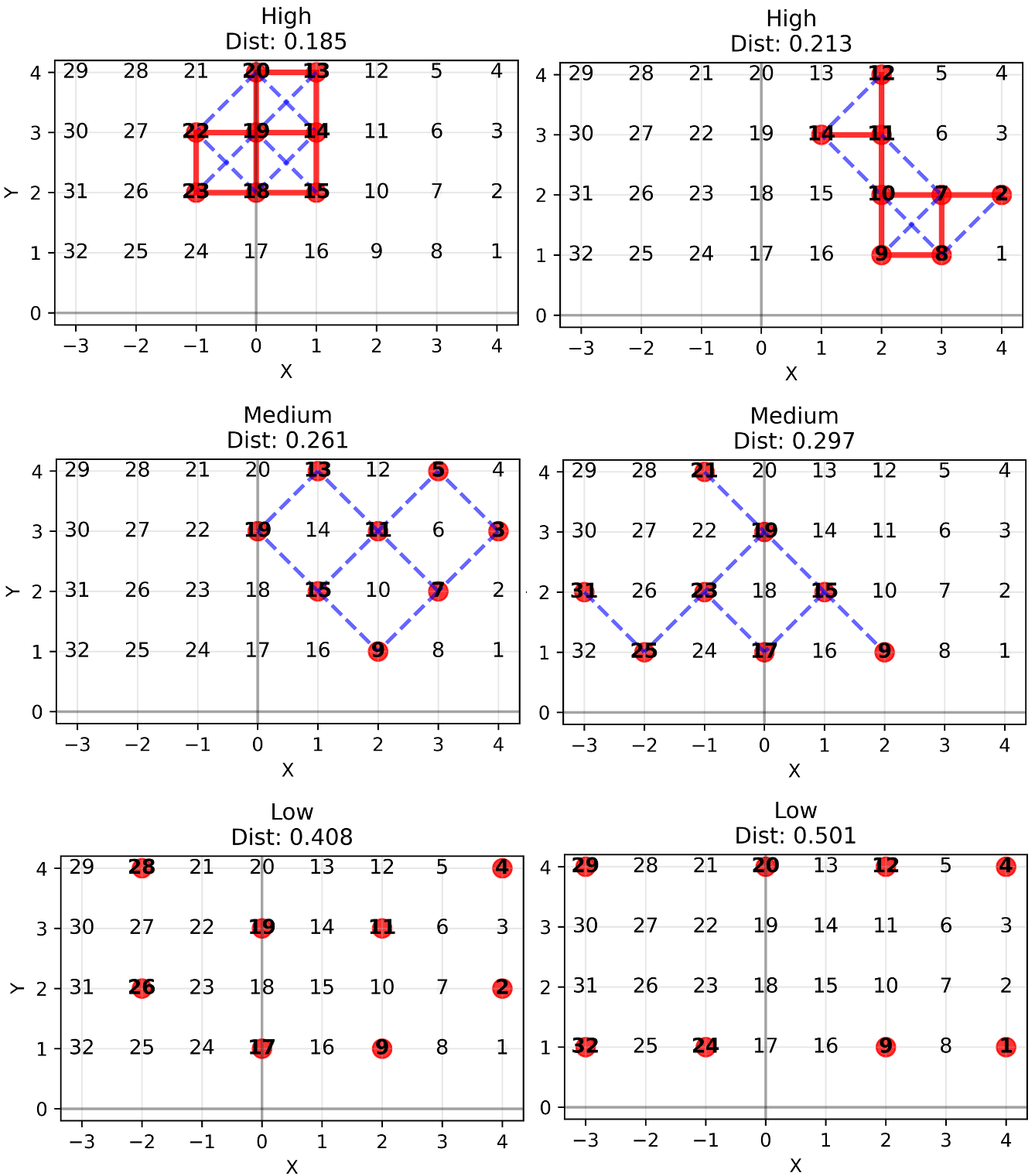}}
\caption{Electrode spatial maps derived from the Maize HD grid showing three levels of spatial sampling density. An eight-electrode configuration is illustrated. High-density maps (top) have shared-edge connectivity (red solid liens). Medium-density maps (middle) have shared-corner connectivity (blue dashed lines) only, and Low-density maps (bottom) consist of spatially isolated electrodes.}
\label{fig_spatial_map}
\end{figure}

\subsection{Thumb Gesture Recognition Models}

CNN and TCN models are widely used in sEMG-based gesture recognition applications \cite{zhong2025deep, zhong2023spatio, meng2022user, ma2024distanet}. To ensure reproducibility, we implemented both CNN and TCN models for system evaluation, with each 250 ms input signal represented as $A \in \mathbb{R}^{N_c \times N_p}$, where $N_c$ denotes the number of channels and $N_p$ denotes the number of temporal samples. 

The architecture of the CNN is illustrated in Fig.~\ref{fig1_system}(i), following a design similar to that in \cite{meng2022user}. The CNN model comprised two convolutional blocks followed by two fully connected (Dense) layers. Each convolutional block consisted of a convolutional (Conv) layer, a batch normalization(BN) layer, a max-pooling (MP) layer, and a dropout layer. The Conv layers used a kernel size of 3, a stride of 1, and zero padding, with the number of filters set to $N_k = 32$ and $64$ for the first and second Conv layers, respectively. BN was used to accelerate convergence, and a leaky ReLU activation function (slope = 0.2) was applied to improve gradient flow. For the MP layers, both the kernel size and stride were set to 3. A dropout rate of 0.2 was used to reduce overfitting. The two Dense layers contained 256 and 6 units, respectively, followed by a softmax layer for six-class gesture classification. The detailed architecture of the TCN model is provided in the \textit{Supplementary Document}.

For both models, the Adam optimizer was used to update network weights, and cross-entropy loss was adopted as the objective function. The initial learning rate was set to 0.001. For each segmented 250 ms signal, the decoding model produced a six-dimensional probability vector, and the gesture class with the highest probability was selected as the final prediction. All experiments were implemented in PyTorch with CUDA acceleration (version 2.0.1 + cu112) on an NVIDIA GeForce RTX 3090 GPU.

\subsection{Performance Evaluation}
A ten-fold cross validation scheme was employed to evaluate the performance of the CNN and TCN models. In each fold, eight blocks were used for training, one block for validation, and one block for testing. The sEMG signals were normalized using the mean and standard deviation values computed from the training set for each channel, applied to the corresponding validation and test sets. This block-wise partitioning ensured that training, validation, and testing samples were temporally disjoint, thereby preventing data leakage. Classification performance was quantified using accuracy, reported as the mean across folds and participants.

\subsection{Electrode Manipulation}
To systematically investigate wrist-based sEMG electrode configurations for thumb gesture recognition, four key factors were examined: muscle region, referencing scheme (monopolar vs.\ bipolar), channel count, and spatial sampling density.

\subsubsection{Selection of Electrode Placement Regions}
Location plays a critical role in the recognition of patterns based on sEMG \cite{wang2024analysis}, particularly for the decoding of thumb movements from the small area of the wrist. Here, to investigate the regional contribution, electrodes located on the extensor and flexor sides of the wrist were evaluated separately for thumb gesture classification.

\subsubsection{Monopolar and Bipolar Wrist sEMG}
The effect of monopolar and bipolar recording configurations was investigated using both Maize and Quattro sensors. The following configurations were evaluated:
\begin{itemize}
    \item M-mono-32: Maize sensor with 32 monopolar channels. 
    \item M-mono-15: Maize sensor with 15 monopolar channels, obtained by randomly selecting eight channels from the extensor grid and seven channels from the flexor grid. For each participant, 100 electrode subsets were randomly sampled and evaluated using ten-fold cross validation (10 folds~$\times$~10 subsets). Ten subjects got 1000 electrode combinations in total.
    \item M-bi-16: Maize sensor with 16 bipolar channels, constructed by differential pairing of non-overlapping odd-even electrode channels, i.e., channels 1-2, 3-4, $\dots$, 31-32. 
    \item Q-bi-15: Quattro sensor with 15 bipolar channels. 
\end{itemize}

\subsubsection{Effect of Channel Counts}

To examine the effect of channel count independent of spatial sampling density, electrode subsets were randomly selected from the Maize and Quattro sensors without imposing spatial connectivity constraints. For the Maize sensor, subsets of 15, 12, 8, 6, and 4 channels were randomly selected from the entire 32-channel grid. For the Quattro sensor, subsets of 12, 8, 6, and 4 channels were randomly selected from the 15 bipolar channels available. For each participant and each channel-count condition, 100 random subsets were generated and evaluated using ten-fold cross-validation (10 folds~$\times$~10 subsets), resulting in 1000 evaluations for ten subjects per configuration. 

\begin{figure*}[ht]
\centerline{\includegraphics[width=\textwidth]{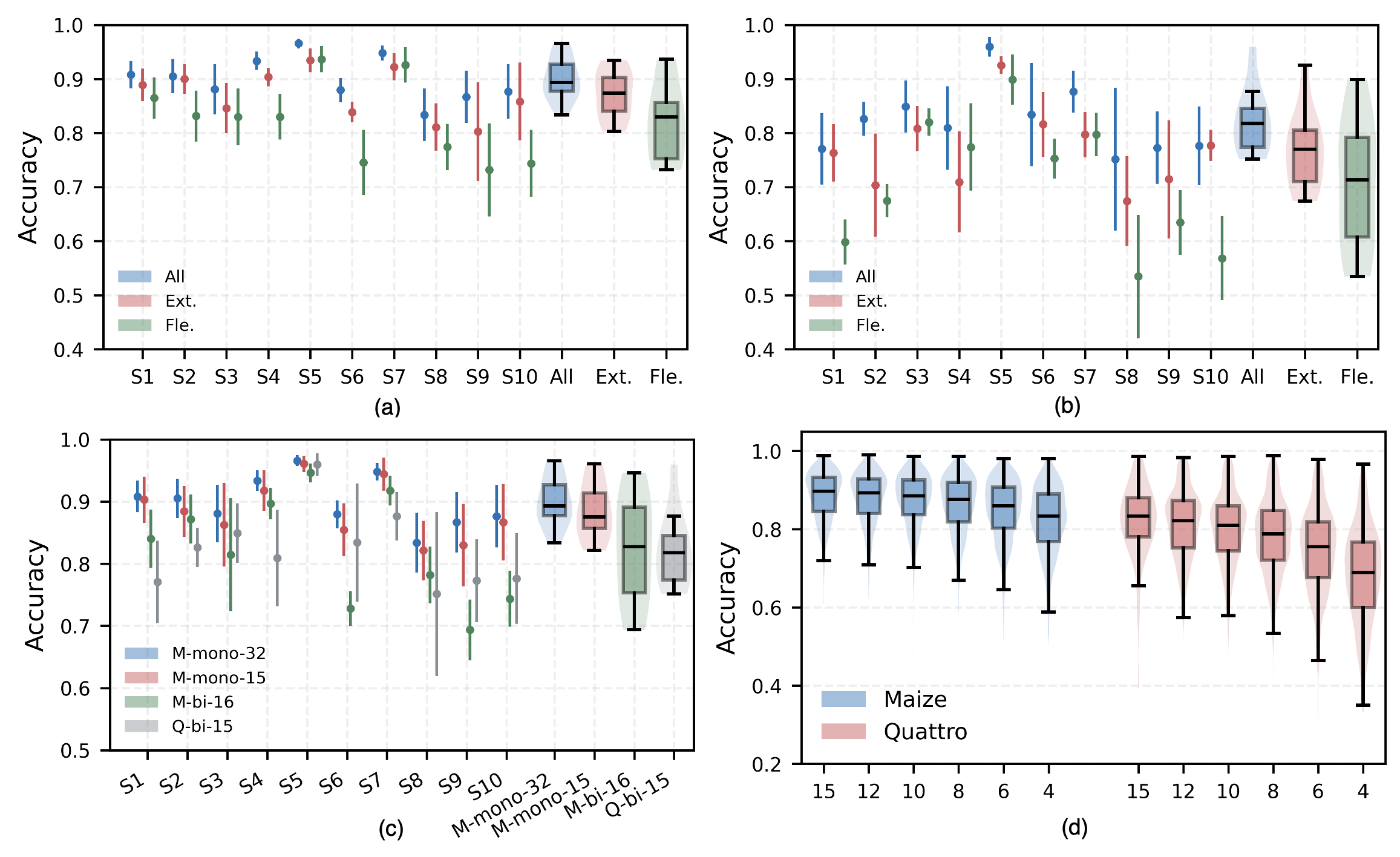}}
\caption{Thumb-movement classification performance under different electrode configurations. 
(a) Performance comparison between extensor (Ext.), flexor (Fle.), and combined electrode placements for Maize sensor.
(b) Corresponding placement comparison for Quattro sensor.
(c) Comparison between monopolar and bipolar configurations using Maize and Quattro sensors across subjects. 
(d) Effect of random channel reduction on classification accuracy for Maize and Quattro sensors.}
\label{acc_ele_configs_cnn}
\end{figure*}

\subsubsection{Electrode Spatial Density Map}
Electrode spatial maps were designed on the basis of the Maize HD grid to investigate the effect of spatial sampling density. Firstly, using the two 4 $\times$ 4 Maize grids (32 channels in total), we constructed a two-dimensional Cartesian coordinate system. Each electrode was assigned a fixed spatial coordination $(x_i, y_i)$ based on its physical configuration on the wrist. 

Secondly, to control spatial density, we defined three electrode maps, i.e., high, medium, and low-density, by constraining inter-electrode connectivity. Taking the eight-channel configuration illustrated in Fig.~\ref{fig_spatial_map} as an example, electrode subsets were randomly selected under the following rules: 
\begin{itemize}
    \item High-density configuration: electrodes were selected so that each electrode shared at least one edge connection with a neighboring electrode, forming compact local clusters.
    \item Medium-density configuration: electrodes were selected with shared-corner connectivity only, without any shared-edge connections.
    \item Low-density configuration: electrodes were spatially isolated, with no shared-edge or shared-corner connections between neighboring electrodes.
\end{itemize}

Third, for each sampled electrode subset, spatial dispersion was quantified by computing all pairwise inter-electrode distances based on their spatial coordinates. Assume $\mathbf{p}_i = (x_i, y_i)$ denotes the spatial coordinate of the $i$-th electrode in the selected subset of $N_c$ electrodes. The pairwise Euclidean distance between electrodes $i$ and $j$ is defined as
\begin{equation}
d_{ij} = \lVert \mathbf{p}_i - \mathbf{p}_j \rVert_2,
\end{equation}
where $i < j$ and $i,j \in \{1, \dots, N_c\}$. The median of all pairwise distances between $N_c$ electrodes was used as a scalar measure of the spatial sampling density, denoted as $\mathrm{Dist}$.
\begin{equation}
\mathrm{Dist} = \mathrm{median}\left( \left\{ d_{ij} \;\middle|\; 1 \le i < j \le N_c \right\} \right).
\end{equation}

To further evaluate the trade-off between decoding performance and spatial sampling density, a figure of merit (FOM) was introduced. The FOM is defined as the ratio between the classification accuracy (ACC) and the spatial distance.
\begin{equation}
\mathrm{FOM} = \frac{\mathrm{ACC}}{\mathrm{Dist}}.
\end{equation}
The FOM provides a compact metric for quantifying the efficiency of electrode configurations by jointly considering the decoding performance and spatial compactness. A higher FOM indicates that a given configuration achieves higher accuracy with a more compact electrode arrangement, whereas a lower FOM reflects either reduced accuracy or increased spatial spread.

ACC and FOM were calculated and analyzed to assess the impact of electrode spatial distribution by sampling subsets of four, six, and eight channels from the Maize HD grids. Given the great number of possible electrode combinations, a random sampling strategy was adopted to ensure computational tractability and generalizability. For each participant, 100 electrode subsets were randomly sampled for each density level (high, medium, and low). During the ten-fold cross-validation procedure, each fold evaluated 10 randomly sampled subsets, resulting in a total of 100 (10 $\times$ 10) evaluated configurations per density level for each participant.

\subsection{Statistical Analysis}
To compare decoding performance across different electrode configurations, a one-way analysis of variance (ANOVA) was performed. Prior to ANOVA, the normality of the data was assessed using the Shapiro--Wilk test. When a significant main effect was observed, \textit{post hoc} pairwise comparisons were conducted using Tukey’s multiple comparison test. A significance threshold of $p < 0.05$ was used for all statistical tests.

\section{Results}
This session presents results of the CNN model, and corresponding results of the TCN model are provided in the \textit{Supplementary Document}.

\begin{figure*}[ht]
\centerline{\includegraphics[width=\textwidth]{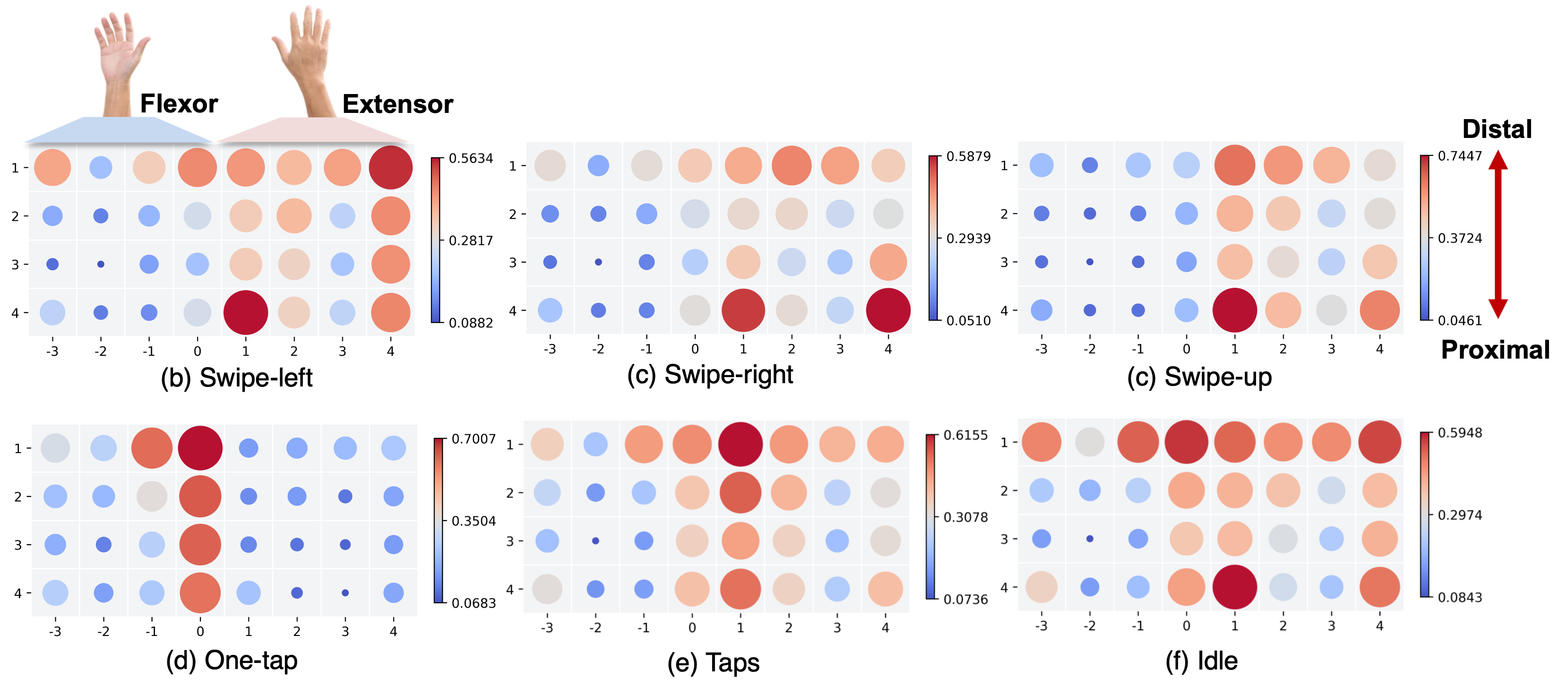}}
\caption{Electrode importance maps for thumb-movement classification using the Maize sensor. (a)--(f) show normalized integrated gradients (IG) attribution distributions obtained from the CNN model for different gestures. Attribution scores were computed with respect to raw sEMG inputs, normalized within each subject, and averaged across subjects. The color scale (blue to red) indicates increasing contribution of each electrode to the CNN model output.}
\label{fig_explain}
\end{figure*}

\subsection{Model Performance and Wrist Placement Regions}
\label{sec:R_msucle area}

Figs.~\ref{acc_ele_configs_cnn}(a) and (b) illustrate the classification performance of different wrist region configurations using the CNN model for the Maize and Quattro sensors, respectively. Three configurations were considered: all electrodes (All), extensor-side electrodes (Ext.), and flexor-side electrodes (Fle.). For the Maize sensor (Fig.~\ref{acc_ele_configs_cnn}(a)), the average accuracies were $0.900 \pm 0.038$, $0.871 \pm 0.044$, and $0.821\pm 0.070$ for the configurations All, Ext. and Fle., respectively. For the Quattro sensor (Fig.~\ref{acc_ele_configs_cnn}(b)), the corresponding accuracies were $0.823 \pm 0.059$, $0.769 \pm 0.070$, and $0.705 \pm 0.115$. Across both sensing systems, the All configuration from both extensor and flexor regions achieved the highest classification accuracy. When considering single-region configurations, electrodes on the extensor side consistently outperformed those on the flexor side.

In addition, the Maize sensor yielded higher performance than the Quattro sensor. For all configurations, Maize achieved higher accuracy than Quattro (mean difference = 0.077, 95\% CI [0.032, 0.122], $p = 0.0019$). This performance gain is likely related to differences in the reference scheme (monopolar vs. bipolar) and channel account, which are further investigated in the following sections.

\subsection{Effect of Reference Scheme: Monopolar vs. Bipolar}
\label{sec:R_mono_bi}

Fig.~\ref{acc_ele_configs_cnn}(c) compares classification performance across monopolar and bipolar reference schemes. The decoding accuracies of M-mono-32, M-mono-15, M-bi-16, and Q-bi-15 were $0.900 \pm 0.038$, $0.885 \pm 0.044$, $0.823 \pm 0.081$, and $0.823 \pm 0.059$, respectively. Overall, monopolar configurations using Maize sensor consistently achieved higher accuracy than bipolar configurations from both Maize and Quattro sensors. Specifically, M-mono-15 significantly outperformed Q-bi-15 (mean difference = 0.062, 95\% CI [0.020, 0.103], $p = 0.0053$). A small accuracy difference was observed between M-mono-32 and M-mono-15 (mean difference = 0.015, 95\% CI [0.005, 0.026], $p = 0.0054$), indicating minimal performance loss when reducing the number of \textit{monopolar} channels. In contrast, a noticeable performance drop was observed for the M-bi-16 configuration compared to M-mono-15 (mean difference = 0.061, 95\% CI [0.020, 0.103], $p = 0.0053$), suggesting that bipolar referencing leads to reduced decoding performance. No statistically significant difference was found between M-bi-16 and Q-bi-15 ($p > 0.9999$).

\subsection{Electrode Reduction}
\label{sec:R_channel reduction}
Motivated by the comparable performance between M-mono-32 and M-mono-15 in Section~\ref{sec:R_mono_bi}, we further evaluated the effect of channel count through progressive electrode reduction. Fig.~\ref{acc_ele_configs_cnn}(d) shows the classification accuracy under random channel selection for both the Maize and Quattro sensors. For the Maize system, the use of 15, 12, 10, 8, 6, and 4 electrodes resulted in accuracies of $0.885 \pm 0.045$, $0.880 \pm 0.048$, $0.875 \pm 0.047$, $0.864 \pm 0.051$, $0.848 \pm 0.054$, and $0.823 \pm 0.060$, respectively. For the Quattro system, the corresponding accuracies were {$0.823 \pm 0.059$, $0.813 \pm 0.063$, $0.801 \pm 0.065$, $0.781 \pm 0.069$, $0.745 \pm 0.073$, and $0.681 \pm 0.083$}.

Classification accuracy decreased progressively as the number of channels retained was reduced. Specifically, no statistically significant difference was observed between 15 and 12 channels (mean difference = 0.010, $p > 0.05$), while significant performance reductions were observed for 10, 8, 6, and 4 channels compared to 15 channels (mean differences = $0.022$, $0.042$, $0.079$, and $0.142$; all $p < 0.05$). 

In contrast, the Maize sensor demonstrated greater robustness to channel reduction, maintaining relatively high accuracy even with fewer channels. Specifically, no statistically significant difference was observed between 15 and 12 channels (mean difference = 0.0049, $p = 0.6757$), while small but statistically significant reductions in performance were observed for 10, 8, 6, and 4 channels compared to 15 channels (mean differences = $0.010$, $0.021$, $0.037$, and $0.063$; all $p < 0.05$).

\subsection{Feature Importance Visualization of Thumb Movements}
\label{sec:channel_importance}

Electrode importance maps were generated using the integrated gradients (IG) method \cite{sundararajan2017axiomatic} to quantify the contribution of Maize electrodes to thumb-movement classification. IG attribution scores were calculated with respect to the raw sEMG input, normalized within each subject, and averaged between subjects to obtain a group-level importance distribution. 

As shown in Fig.~\ref{fig_explain}, common spatial trends were observed in gestures. Electrodes located in the distal region of the wrist exhibited higher attribution values than those in the proximal region. In addition, a clear radial–ulnar asymmetry was observed, with a greater importance on the radial side. Gesture-specific differences were also evident between the flexor and extensor regions. For the \textit{one-tap} gesture, the higher importance values were concentrated on the flexor aspect. In contrast, for \textit{swipe-up}, \textit{swipe-left}, \textit{swipe-right}, and \textit{taps} gestures, electrodes on the extensor side consistently exhibited the highest attribution values.

\subsection{The Effect of Electrode Spatial Density}
\label{sec:R_density}

\begin{figure*}[ht]
\centerline{\includegraphics[width=\textwidth]{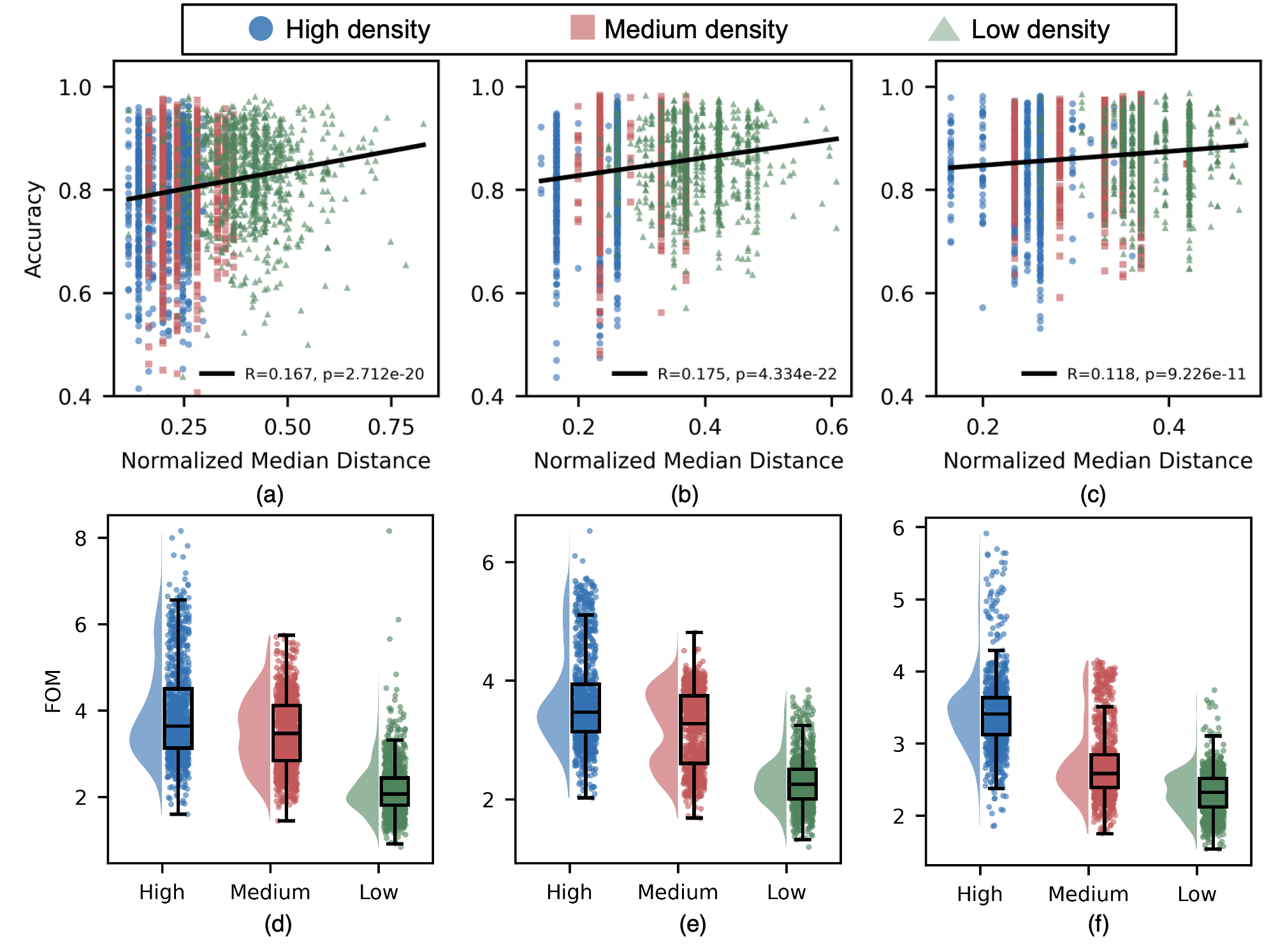}}
\caption{Effect of electrode spatial density under four-, six-, and eight-channel configurations using the Maize sensor. (a)--(c) relationship between classification accuracy and spatial distance ($\mathrm{Dist}$). (d)--(f) The relationship between the figure of merit (FOM) and electrode density. Each point represents one randomly sampled electrode configuration.}

\label{density}
\end{figure*}

Figs.~\ref{density}(a)--(c) show the relationship between classification accuracy and spatial distance ($\mathrm{Dist}$) for four-, six-, and eight-channel configurations, respectively. For a certain number of channels, $\mathrm{Dist}$ serves as a measurement of electrode spacing and inversely reflects spatial density. In all configurations, a subtle but statistically significant positive correlation was observed between $\mathrm{Dist}$ and classification accuracy. Linear regression yielded correlation coefficients of $R = 0.167$ ($p = 2.712 \times 10^{-20}$), $R = 0.175$ ($p = 4.334 \times 10^{-22}$) and $R = 0.118$ ($p = 9.226 \times 10^{-11}$) for four-, six-, and eight-channel configurations, respectively. This indicates that slightly larger electrode spacing (i.e., lower density) is associated with improved decoding performance. 

Figs.~\ref{density}(d)--(f) present the relationship between the FOM and electrode density. In contrast to accuracy, the FOM peaked at higher electrode densities, indicating improved efficiency under more compact spatial configurations. 

\section{Discussion}
This study developed a multi-modal experimental platform that enables synchronized acquisition of wrist sEMG signals and video recordings of hand movements. Using this platform, we recorded wrist sEMG signals with both HD (Maize) and LD (Quattro) sensors. We developed both CNN and TCN architectures for verifying different electrode configurations by decoding thumb movements. The decoding performances reached $0.900 \pm 0.038$ (Maize) and $0.823 \pm 0.059$ (Quattro) with the CNN model, and $0.826 \pm 0.063$ (Maize) and $0.736 \pm 0.075$ (Quattro) with the TCN model. With the purpose of investigating wrist-based configurations for sEMG electrodes, we conducted a systematic analysis of key sEMG sensing factors, including wrist muscle region, sEMG reference scheme, electrode count, and spatial sampling density.

\subsection{Flexor and Extensor Muscle Contributions}

The results in Section~\ref{sec:R_msucle area} show that combining signals from both the flexor and the extensor regions produces the highest decoding performance, while extensor-only configurations consistently outperform flexor-only configurations. In contrast, flexor-based decoding exhibits greater inter-subject variability. The performance advantage of the muscle groups on the extensor-side can be explained from an anatomical perspective. Wrist sEMG predominantly captures the activity of extrinsic thumb muscles whose tendons traverse the wrist joint. On the extensor side, key muscles associated with thumb gestures, such as abductor pollicis longus (APL), extensor pollicis longus (EPL), and extensor pollicis brevis (EPB), follow relatively distinct anatomical pathways \cite{emerson1996anatomy}. This spatial separation improves muscle activation separability and enhances the spatial selectivity of recorded sEMG signals. In contrast, the flexor side is dominated by the pollicis longus (FPL) flexor, whose tendon passes through a crowded anatomical region together with other finger flexors \cite{emerson1996anatomy}. This arrangement increases inter-muscle overlap and crosstalk, reducing discriminative information and leading to less stable decoding performance. 

From a system design perspective, these findings suggest that incorporating electrodes over both muscle regions is beneficial for maximizing information capture, while prioritizing extensor-side coverage can improve robustness when channel count is limited.

In addition, the Maize sensor consistently outperformed the Quattro sensor in all configurations. This performance gap suggests that factors beyond the muscle region, including the reference scheme (monopolar vs.~bipolar), channel count, and spatial density, play a critical role in determining decoding performance. To isolate these effects, we further investigated each factor, including referencing scheme, channel count, and spatial density, to analyze their individual contributions.

\subsection{Monopolar and Bipolar} 

Fig.~\ref{acc_ele_configs_cnn} (c) shows that the monopolar reference scheme consistently outperformed the bipolar recordings. Two factors may contribute to this performance difference. First, monopolar sEMG preserves the original muscle activation signals without inter-channel differencing, thereby retaining higher spatial information content. In contrast, bipolar recordings emphasize local differences between adjacent electrodes, which can attenuate broader activation patterns. As a result, monopolar signals provide richer spatial features and improved separability of muscle activation patterns.

Second, monopolar configurations exhibited lower inter-subject variability compared to bipolar configurations. As shown in Fig.~\ref{acc_ele_configs_cnn}(c), bipolar recordings demonstrated greater variance in performance between subjects. This may be attributed to the higher sensitivity of bipolar measurements to electrode placement. Accurate bipolar recordings require proper alignment of electrode pairs with the underlying muscle fibers, and misalignment can degrade signal quality. In contrast, monopolar sEMG recordings with high spatial resolution are less dependent on precise electrode pairing, making them more robust and stable in classification performance between subjects.

\subsection{Effect of Electrode Count}

Fig.~\ref{acc_ele_configs_cnn}(a) shows that increasing the number of sEMG channels improves decoding performance, particularly for the Quattro sensor (Section~\ref{sec:R_channel reduction}). However, results from the Maize sensor indicate a performance saturation effect, where increasing the channel count beyond a moderate level yields limited improvement. In particular, eight channel configurations achieved performance comparable to the 15-channel configuration, despite a statistically significant difference (mean difference = 0.021, $p = 0.0049$). This suggests that the absolute performance gain from additional channels is small, even though the difference is statistically significant. From a system design perspective, increasing the number of channels also leads to higher computational and hardware costs. A reduced channel configuration (e.g., eight channels) can provide a favorable trade-off between decoding accuracy and system complexity.

\subsection{Model Explainability}

The electrode importance patterns revealed by the IG analysis provide insights into the physiological origins of wrist sEMG signals during thumb movements. The spatial distribution of important electrodes was consistent with previously reported activation patterns of far-field electric potentials recorded from wrist-mounted sensors \cite{guerra2022far}. In particular, the observed radial–ulnar and proximal–distal asymmetries in the attribution maps (Fig.~\ref{fig_explain}) align with the spatial distribution of the root mean square activity (RMS) and the decoded motor neuron potentials reported in a previous study \cite{guerra2022far}. These results suggest that the explainability analysis captures physiologically meaningful signal structures underlying wrist sEMG recordings.

The explainability results further highlight the dominant role of extensor-side signals in thumb-movement decoding. For most gestures, electrodes located in the extensor region exhibited higher attribution scores, indicating that these channels contribute more to CNN predictions. This observation is consistent with the classification results presented in Fig.~\ref{acc_ele_configs_cnn}(a), where the models that use electrodes on the extensor-side achieved a higher decoding accuracy than those that use flexor-side signals. From a practical perspective, these findings suggest that electrode placement strategies for wrist-worn sEMG interfaces should prioritize coverage of the extensor region, especially when the number of available electrodes is limited. Furthermore, the spatial importance maps in Fig.~\ref{fig_explain} provide a potential guide for optimizing electrode placement in compact designs worn on the wrist. 

\subsection{Effects of Electrode Density}

The results show a weak but consistent positive correlation between classification accuracy and spatial distance ($\mathrm{Dist}$), indicating that lower electrode density (i.e., larger spacing) slightly improves decoding performance. This trend suggests that increased spatial coverage allows the sensing system to capture more diverse activation patterns. However, the effect size remains small across all channel-count configurations, indicating that decoding performance is only weakly dependent on electrode spacing at the wrist. In contrast, the FOM exhibits an opposite trend, with relatively higher values achieved at higher electrode densities. These findings reveal a trade-off between spatial coverage and electrode efficiency. Although lower electrode density (larger spacing) may improve classification accuracy performance by reducing spatial redundancy and capturing broader muscle activity, higher density configurations enhance efficiency by concentrating electrodes within a compact area with a slight reduction in classification accuracy. 

This behavior differs from previous studies on the forearm \cite{he2018electrode}, where increasing electrode sampling density over a larger acquisition area leads to significant improvements in decoding performance. In contrast, the wrist represents a compact sensing region with closely packed tendons and overlapping muscle contributions, limiting the benefit of spatial sampling of the electrode.
From a system design perspective, these results suggest that effective wrist-worn sEMG systems can be achieved using a compact set of strategically placed electrodes, rather than relying on lower density sensing configurations.

These observations also motivate the exploration of alternative sensing modalities that can overcome the spatial limitations inherent in wrist sEMG. In particular, the measurement of magnetic muscle activity, or magnetomyography (MMG), offers a promising complementary approach \cite{zuo2020miniaturized, yun2024magnetomyography}. Unlike sEMG, which relies on differential voltage measurements and is constrained by electrode placement and tissue conductivity, MMG captures the magnetic fields generated by muscle electrical activity and can be measured without contact at the skin surface. This enables inherently higher spatial resolution and reduced sensitivity to electrode configuration and skin-electrode interface variability. Furthermore, MMG signals are less affected by volume conduction, allowing improved separation of closely spaced muscle sources, which is particularly relevant in anatomically compact regions such as the wrist. As a result, MMG has the potential to alleviate the trade-off between spatial coverage and electrode density observed in sEMG, enabling compact sensing systems with enhanced decoding performance. Future work should therefore investigate hybrid or standalone MMG-based wrist interfaces to further advance high-fidelity, wearable human–machine interaction.

\subsection{Implications for Wrist-Worn sEMG Designs}

Recent studies have compared sEMG signal characteristics and decoding performance between wrist and forearm recordings, demonstrating the feasibility and potential of wrist-worn sEMG for human-machine interaction \cite{kaifosh2025generic}. From Section.~\ref{sec:R_msucle area},~\ref{sec:R_mono_bi},~\ref{sec:R_channel reduction}, and~\ref{sec:R_density}, we quantitatively examined the configurations of wrist sEMG sensors in multiple design dimensions, which has direct implications on wearable form factors where the sensing area and power consumption are limited.

In addition, multi-modal sensor fusion represents a promising direction for further improving robustness in wrist-worn interfaces; for example, combining sEMG with inertial measurements (IMU) \cite{jiang2017feasibility, krasoulis2017improved} improves performance. Other studies have developed wrist-based gesture recognition applications based on muscle impedance signals \cite{zhu2025ei} and mechanomyography \cite{zhu2025ei}. Integrating such complementary modalities may enrich the data set and improve the development of models for robust and generative applications.

Hardware design inevitably involves a trade-off between user comfort and sensing capability. High-resolution and large-area electrode arrays improve decoding performance but increase system complexity, size, and power requirements \cite{yang2025non, simpetru2024learning}. Our electrode importance visualization and FOM results would provide guidance for electrode placement and layout for wrist systems. Furthermore, recent advances in wireless electronics and simplified electrode layouts \cite{kim2025simplified, cao2024finger} suggest that combining optimized electrode placement strategies with lightweight hardware design may offer a viable path toward practical wrist-worn sEMG interfaces. 

\bibliographystyle{IEEEtran}
\bibliography{ref.bib}
\vfill
\end{document}